\documentclass{amsart}

\usepackage{amssymb,latexsym,amsfonts,amsmath}
\usepackage{graphicx}
\include{diagrams}
\usepackage{eso-pic}
\usepackage{graphicx}
\usepackage{color}
\usepackage{type1cm}

\usepackage{amssymb,latexsym,amsfonts,amsmath}
\usepackage{graphicx}

\newtheorem{theorem}{Theorem}[section]

\newtheorem{proposition}[theorem]{Proposition}

\theoremstyle{definition}

\theoremstyle{remark}

\numberwithin{equation}{section}

\newcommand{\R}{{\mathbb{R}}}

\newcommand{\N}{{\mathbb{N}}}
\newcommand{\T}{{\mathbb T}}

\renewcommand{\T}{\mathbf{T}}

\begin{document}
\begin{abstract}
The paper~\cite{TF19} proposes a data-driven control technique for single-input single-output feedback linearizable systems with unknown control gain by relying on a persistency of excitation assumption. This note extends those results by showing that persistency of excitation is not necessary. We refer the readers to the papers~\cite{TMGA17,TF19} for more background and motivation for the technical results in this note. Conceptually, the results in this note were greatly inspired by the work of Fliess and Join on intelligent PID controllers, e.g.,~\cite{FJ09}. Technically, we were inspired by the work of Nesic and co-workers on observer and controller design based on approximate models~\cite{AN04,NT04} and by the work of Astolfi and Ortega on Immersion and Invariance~\cite{AO03}.
\end{abstract}

\title[Data-driven control]{A note on data-driven control for\\SISO feedback linearizable systems\\without persistency of excitation}

\author[Paulo Tabuada]{Paulo Tabuada}
\address{Department of Electrical and Computer Engineering\\ 
University of California at Los Angeles\\
Los Angeles, CA 90095-1594, USA}
\urladdr{http://www.ee.ucla.edu/$\sim$tabuada}
\email{tabuada@ee.ucla.edu}

\author[Lucas Fraile]{Lucas Fraile}
\address{Department of Electrical and Computer Engineering\\ 
University of California at Los Angeles\\
Los Angeles, CA 90095-1594, USA}
\email{lfrailev@ucla.edu}

\maketitle

\section{Notation}
The natural numbers, including zero, are denoted by $\N$, the real numbers by $\R$, the non-negative real numbers by $\R_0^+$, and the positive real numbers by $\R^+$. If $c:\R\to \R^n$ is a function of time, we denote its first time derivative by $\dot{c}$. When higher time derivatives are required, we use the notation $c^{(k)}$ defined by the recursion $c^{(1)}=\dot{c}$ and $c^{(k+1)}=\left(c^{(k)}\right)^{(1)}$. The Lie derivative of a function $h:\R^n\to \R$ along a vector field $f:\R^n\to\R^n$, given by $\frac{\partial h}{\partial x}f$, is denoted by $L_fh$. 

Consider the function $f:\R_0^+\times \mathcal{S}\to \R^n$, $\mathcal{S}\subseteq \R^n$. We will use the notation $f(t,x)=O_x(T)$ to denote the existence of constants $M,T\in \R^+$ so that for all $t\in [0,T]$ and $x\in \mathcal{S}$ we have $\Vert f(t,x)\Vert \le MT\Vert x\Vert$. We will use the following rules that apply to this notation where the equalities below mean that the left-hand side can be replaced by the right-hand side:
$$O_x(T^2)=O_x(T),\quad \left(O_x(T)\right)^2=O_{x^2}(T^2),\quad TO_x(T)=O_x(T^2),\quad g(x)O_x(T)=O_x(T).$$
The subscript $x^2$ in $O_{x^2}(T^2)$ indicates we are squaring the norm, i.e., $O_{x^2}(T^2)$ denotes the upper bound $MT^2\Vert x\Vert^2$. Moreover, the function  $g$ has bounded norm, i.e., there exists $b\in \R^+$ so that $\Vert g(x)\Vert\le b$ for all $x\in \mathcal{S}$.

\section{Models}
\label{Sec:Models}
We consider an unknown single-input single-output nonlinear system described by:
\begin{eqnarray}
\label{UnknownSystem1a}
\dot{x}&=&f(x)+g(x)u\\
\label{UnknownSystem1b}
y&=&h(x)+d,
\end{eqnarray}
where $f:\R^n\to\R^n$, $g:\R^n\to\R^n$, and $h:\R^n\to \R$ are smooth functions and  we denote by $y\in \R$, $x\in \R^n$, $u\in \R$, $d\in\R$, the output, state, input, and measurement noise, respectively. We make the assumption that the output function $h$ has relative degree $n$, i.e., this system is feedback linearizable. This means that $L_gL_f^ih(x)=0$ for $i=0,\hdots,n-2$ and $L_gL_f^{n-1}h(x)\ne 0$ for all $x\in \R^n$. Since the function $L_gL_f^{n-1}h$ is continuous and never zero, its sign is constant. We will assume the sign of $L_gL_f^{n-1}h$ to be know and, without loss of generality, take it to be positive. Moreover, we assume the knowledge of two constants $\underline{\beta}, \overline{\beta}\in \R^+$ such that:
\begin{equation}
\label{AssumptionFL}
\underline{\beta}\le L_gL_f^{n-1}h(x)\le \overline{\beta},
\end{equation}
for all $x\in \R^n$. Knowledge of the sign of $L_gL_f^{n-1}h$ is not a strong assumption beyond $L_gL_f^{n-1}h\ne 0$. A simple input/output measurement experiment can be performed to infer the sign of $L_gL_f^{n-1}h$. Knowledge of the constants $\underline{\beta}$ and $\overline{\beta}$ is a stronger assumption that can be justified whenever some structural information about the system is available. For example, if a point mass with mass $m$ moves in a one dimensional line by the action of a force $u$, we have the model $m x^{(2)}=u$ where $x$ denotes the mass position that is measured by a sensor, i.e., $y=x$. In this case we have $L_gL_f^{n-1}h=1/m$ and thus knowledge of the bounds $\underline{\beta}$ and $\overline{\beta}$ requires some prior knowledge about the range of possible mass values.

With the objective of presenting the results in its most understandable form, we assume $n=2$ throughout this note, although all the results hold for arbitrary $n\in \N$. This will enable us to preform all the necessary computations explicitly and without the need for distracting bookkeeping. 

Invoking the feedback linearizability assumption, we can rewrite the unknown dynamics in the coordinates $(z_1,z_2)=\Psi(x)=(h(x),L_fh(x))$:
\begin{eqnarray}
\label{UnknownSystem2a}
\dot{z}_1&=&z_2\\
\label{UnknownSystem2b}
\dot{z}_2&=&\alpha(z)+\beta(z)u\\
\label{UnknownSystem2c}
y & = & z_1,
\end{eqnarray}
where $\alpha=L_f^2h\circ\Psi^{-1}$ and $\beta=L_gL_fh\circ\Psi^{-1}$. We note that $f$, $g$, and $h$ are unknown and thus so are $\alpha$ and $\beta$. This form of the dynamics has the advantage of using the two scalar valued functions $\alpha$ and $\beta$ to describe the full dynamics, independently of the value of $n$. We exploit this observation by learning the values of these functions online.

System~\eqref{UnknownSystem2a}-\eqref{UnknownSystem2c} will be controlled using piecewise constant inputs for a sampling time $T\in \R^+$. This means that inputs $u:\R_0^+ \to \R$ satisfy the following equality for all $k\in \N$:
$$u(kT)=u(kT+\tau),\qquad \forall \tau\in [0,T[.$$ It will be convenient to use $u$ to denote an input only defined on $[0,T[$. Since the curve $u$ is constant on the interval $[0,T[$, we identify it with the corresponding element of $\R$.

The solution of~\eqref{UnknownSystem2a}-\eqref{UnknownSystem2b} is denoted by $F_t^e(z,u)=(F_{t,1}^e(z,u),F_{t,2}^e(z,u))$, for $t\in [0,T[$, and satisfies $F_0^e(z,u)=z$. The superscript ``$e$'' denotes that this is an exact solution. In the next section we discuss approximate solutions.

\section{Approximate models}
In this section we develop an approximate solution of~\eqref{UnknownSystem2a}-\eqref{UnknownSystem2b} based on the well known Taylor's theorem that we now recall.

\begin{theorem}
Let $c:\mathcal{I}\to \R^n$ be an $n$ times differentiable function where $\mathcal{I}\subseteq \R$ is an open and connected set. For any $t,\tau\in \mathcal{I}$ such that $\tau+t\in \mathcal{I}$ we have:
$$c(\tau+t)=c(\tau)+c^{(1)}(\tau)t+c^{(2)}(\tau)\frac{t^2}{2}+\hdots+c^{(n-1)}(\tau)\frac{t^{n-1}}{(n-1)!}+c^{(n)}(\tau')\frac{t^{n}}{n!},$$
for some $\tau'\in [\tau,\tau+t]$.
\end{theorem}

Applying this result to $F_{\tau+t,1}^e$ we obtain:
$$F_{\tau+t,1}^e(z,u)=F_{\tau,1}^e(z,u)+\left(F_{\tau,1}^e\right)^{(1)}(z,u)t+\left(F_{\tau,1}^e\right)^{(2)}(z,u)\frac{t^2}{2}+\left(F_{\tau',1}^e\right)^{(3)}(z,u)\frac{t^3}{3!}.$$
If we only retain the first three terms we obtain an approximate solution with an  approximation error given by the magnitude of the (neglected) fourth term. The following result provides a bound for the approximation error in a form useful for the results derived in this note.

\begin{proposition}
\label{Prop:BoundTaylor}
Let $\mathcal{D}\subset\R^3$ be a compact set. Then, there exist $T\in \R^+$ and $M\in \R^+$ such that:
\begin{equation}
\label{BoundInProp}
\left\Vert \left(F_{\tau',1}^e\right)^{(3)}(z,u)\frac{t^3}{3!}\right\Vert\le M{T}^3\Vert (z,u-u_0)\Vert,
\end{equation}
for all $(z,u)\in \mathcal{D}$, all $t,\tau'\in [0,T]$, and where $u_0=-\alpha(0)\beta^{-1}(0)$.
\end{proposition}

Using the $O$ notation, this result states that:
$$\left(F_{\tau',1}^e\right)^{(3)}(z,u)\frac{t^3}{3!}=O_{(z,u-u_0)}(T^3).$$

\begin{proof}
Since~\eqref{UnknownSystem2a}-\eqref{UnknownSystem2b} is a smooth differential equation (recall that inputs are constant), solutions exist for all $\tau\in [0,T_{z,u}[$ where $[0,T_{z,u}[$ is the maximal interval for which the solution $F_{\tau,1}^e(z,u)$ exists. The function $(z,u)\mapsto T_{z,u}$ is lower semi-continuous and, given that $(z,u)$ belongs to the compact set $\mathcal{D}$, it achieves its minimum on $\mathcal{D}$. Let $T\in \R^+$ be smaller than $\min_{(z,u)\in \mathcal{S}}T_{z,u}$. By definition of $T$, for any $(x,u)\in \mathcal{D}$ solutions exist on the interval $[0,T]$. Consider now the function $\left(F_{\tau',1}^e\right)^{(3)}$ that is continuously differentiable, by assumption, and thus Lipschitz continuous on $\mathcal{D}\times \{\tau'\}$ for each fixed $\tau'\in [0,T]$. Hence, we have:
\begin{equation}
\label{Eq:Lips}
\left\Vert \left(F_{\tau',1}^e\right)^{(3)}(z,u)-\left(F_{\tau',1}^e\right)^{(3)}(z',u')\right\Vert\le L(\tau')\Vert (z,u)-(z',u')\Vert,
\end{equation}
forall $(z,u),(z',u')\in \mathcal{D}$ and all $\tau'\in [0,T]$. Noting that, according to~\eqref{UnknownSystem2a}-\eqref{UnknownSystem2b},  $F_{\tau'}^e(0,u_0)=0$ for $u_0=-\alpha(0)\beta^{-1}(0)$ and all $\tau' \in [0,T]$, we conclude that $\left(F_{\tau',1}^e\right)^{(3)}(0,u_0)=0$. Using this equality in~\eqref{Eq:Lips} we obtain:
$$\left\Vert \left(F_{\tau',1}^e\right)^{(3)}(z,u)\right\Vert\le L(\tau')\Vert (z,u)-(0,u_0)\Vert=L(\tau')\Vert (z,u-u_0)\Vert,$$
by setting $z'=0$ and $u'=-u_0$. If we now take $M=\frac{1}{3!}\max_{\tau' \in[0,T]}L(\tau')$ we obtain the desired inequality. Note that $M$ is well defined since $L$ is continuous and $[0,T]$ compact.
\end{proof}

Based on Proposition~\ref{Prop:BoundTaylor} we can write the solution $F_t^e$ of~\eqref{UnknownSystem2a}-\eqref{UnknownSystem2b}, valid for all $t\in [0,T[$, as:
\begin{eqnarray}
\label{ModelOrder3a}
F_{t,1}^e(z,u)&=&z_1+z_2t+(\alpha(0)+\beta(0)u)\frac{t^2}{2}+O_{(z,u-u_0)}(T^3)\\
\label{ModelOrder3b}
F_{t,2}^e(z,u)&=&z_2+(\alpha(0)+\beta(0)u)t+O_{(z,u-u_0)}(T^2).
\end{eqnarray}
Note that $\alpha$ and $\beta$ can be treated as constants since it follows from Proposition~\ref{Prop:BoundTaylor} that:
\begin{equation}
\label{Constants}
\alpha(t)=\alpha(0)+O_{(z,u-u_0)}(T),\qquad \beta(t)=\beta(0)+O_{(z,u-u_0)}(T).
\end{equation}
Hence, for simplicity of notation, we drop the argument of $\alpha$ and $\beta$ and write:
\begin{eqnarray}
\label{ModelOrder23a}
F_{t,1}^e(z,u)&=&z_1+z_2t+(\alpha+\beta u)\frac{t^2}{2}+O_{(z,u-u_0)}(T^3)\\
\label{ModelOrder23b}
F_{t,2}^e(z,u)&=&z_2+(\alpha+\beta u)t+O_{(z,u-u_0)}(T^2).
\end{eqnarray}
By setting\footnote{Although $t\in [0,T[$, solutions are not altered by changing the input on a zero measure set.} $t$ equal to $T$, the previous model provides a \emph{family} of discrete-time \emph{approximate} models indexed by $T$:
\begin{eqnarray}
\label{ModelOrder33a}
z_1(k+1)&=&F_{T,1}^a(z,u)=z_1(k)+z_2(k)T+(\alpha+\beta u(k))\frac{T^2}{2}\\
\label{ModelOrder33b}
z_2(k+1)&=&F_{T,2}^a(z,u)=z_2(k)+(\alpha+\beta u(k))T,
\end{eqnarray}
where $z(k)$ denotes the value of $z$ at time $kT$, $k\in \N$.

When we need to refer to the exact solution of~\eqref{UnknownSystem2a}-\eqref{UnknownSystem2b} we use the notation $F_T^e(z,u)$ while the notation $F_T^a(z,u)$, or simply $z(k+1)$, is reserved for the solution of the approximate discrete-time model~\eqref{ModelOrder33a}-\eqref{ModelOrder33b}.

\section{State estimation}
\label{Sec:StateEstimation}
While control inputs only change every $T$ units of time, measurements will be made more frequently. More precisely, they will be made every $T/\rho$ units of time for some $\rho\in \{n+1,n+2,\hdots\}$. To emphasize this fact, we index the sampled output using the variable $\ell\in \N$, i.e., we write $y(\ell)$ to denote $y(\ell T/\rho)$ and write $y(k)$ to denote $y(k T)$. Therefore, the equality $y(k)=y(\ell)$ only holds when $\ell=\rho k$. Moreover, $\rho$ samples of the output $y$ are acquired while the input remains constant. Specifically, all the samples $y(\ell+1),y(\ell+2),\hdots, y(\ell+\rho)$ for $\ell=\rho k$ correspond to the same input $u(k)$. Although at time $\ell+\rho=\rho(k+1)$ the input $u(k+1)$ is applied, the value of the measurement $y(k+1)$ is defined by the input $u(k)$ and not by the input $u(k+1)$.

While the input is being held constant, i.e., for $\ell\in \{\rho k,\hdots,\rho(k+1)-1\}$ we have the following family of approximate models:
\begin{eqnarray}
\label{ApproxModel3States1}
z_1(\ell+1) &=& z_1(\ell)+z_2(\ell)T+z_3(\ell)\frac{T^2}{2}\\
\label{ApproxModel3States2}
z_2(\ell+1) &=& z_2(\ell)+z_3(\ell)T\\
\label{ApproxModel3States3}
z_3(\ell+1)&=&z_3(\ell),
\end{eqnarray}
where $z_3(\ell)=\alpha+\beta u(\rho k)$. Being a linear model, it can be written in the form:
$$z(\ell+1)=Az(\ell),\qquad z_1(\ell)=y(\ell)=Cz(\ell).$$ 
It can be easily checked that $A$ is invertible and we thus denote by $\mathcal{O}$ the observability matrix for the pair $(A^{-1},C)$ which allows us to write for $\ell=\rho k$:
$$\begin{bmatrix}y(\ell)\\y(\ell-1)\\\vdots\\y(\ell-\rho+1)\end{bmatrix}=Y(\ell)=\mathcal{O}z(\ell).$$
An estimate $\widehat{z}(k)$ of the state vector $z(k)$ can then be obtained by solving this equation via least-squares:
\begin{equation}
\label{StateEstimateFinal}
\widehat{z}(k)=(\mathcal{O}^{T}\mathcal{O})^{-1}\mathcal{O}^T Y(\rho k).
\end{equation}
The following simple computation:
\begin{eqnarray}
&&Y=\mathcal{O}z+O_{(z,u-u_0)} \left((T/\rho)^3\right)=\mathcal{O}z+O_{(z,u-u_0)} \left(T^3\right)\notag\\
&\implies&\mathcal{O}^TY=\mathcal{O}^T\mathcal{O}z+\mathcal{O}^TO_{(z,u-u_0)} (T^3)\notag\\
&=&\mathcal{O}^TY=\mathcal{O}^T\mathcal{O}z+O_{(z,u-u_0)} \left(\begin{bmatrix}T^3\\T^4\\T^5\end{bmatrix}\right)\notag\\
&\implies&(\mathcal{O}^T\mathcal{O})^{-1}\mathcal{O}^TY=z+(\mathcal{O}^T\mathcal{O})^{-1}O_{(z,u-u_0)} \left(\begin{bmatrix}T^3\\T^4\\T^5\end{bmatrix}\right)\notag\\
&\implies&(\mathcal{O}^T\mathcal{O})^{-1}\mathcal{O}^TY=z+O_{(z,u-u_0)} \left(\begin{bmatrix}T^3\\T^2\\T\end{bmatrix}\right)\notag,
\end{eqnarray}
shows that $\widehat{z}=z+O_{(z,u-u_0)} (T)$. If we introduce the estimation error $e_z$, defined by $e_z=z-\widehat{z}$, it follows that:
\begin{equation}
\label{StateEstimationError}
e_z=O_{(z,u-u_0)}(T).
\end{equation}
The control scheme proposed in Section~\ref{Sec:ControllerDesign} only depends on the preceding equality. Hence, we can replace least-squares estimation with any other estimation technique leading to~\eqref{StateEstimationError}. In particular, the parameter $\rho$ is not relevant to the theoretical analysis although it will play an important role in mitigating the effect of sensor noise: longer values of $\rho$ ``average out'' the effect of noise. It is also possible to use the same sampling rate for estimation and control. However, in that case we need to jointly estimate the state and the parameters $\alpha$ and $\beta$ whereas the current approach lets us treat these two problems separately.

\section{Parameter estimation}
\label{SecParameter}
We now consider the problem of estimating the parameters $\alpha$ and $\beta$. Although we are not using an observer based on Immersion and Invariance~\cite{IIBook}, the proof technique used in the main results is directly inspired by the analysis in~\cite{IIBook}. Hence, already in this section we start to see some similarities as we introduce a simple Luenberger observer and set the stage for its Lyapunov analysis. To simplify the presentation we will assume the state $z$ to be known. The proposed observer is given by:
$$\widehat{\pi}(k+1)=\widehat{\pi}(k)+\gamma(k)\Phi(k)\left(z_2(k+1)-z_2(k)-T  \Phi^T(k)\widehat{\pi}(k)\right),$$ 
where $\gamma(k)\in \left[\underline{\gamma},\overline{\gamma}\right]$, $\underline{\gamma},\overline{\gamma}\in \R^+$, is a time-varying observer gain, $\widehat{\pi}$ is the estimate of $\pi$:
$$\widehat{\pi}=\begin{bmatrix}\widehat{\alpha}\\\widehat{\beta}\end{bmatrix},\qquad {\pi}=\begin{bmatrix}{\alpha}\\{\beta}\end{bmatrix},$$
and $\Phi$ is defined by:
$$\Phi(k)=\begin{bmatrix}1\\u(k)\end{bmatrix}.$$
If we define the observer error as: 
\begin{equation}
\label{ParameterError}
e_\pi=\pi-\widehat{\pi},
\end{equation} 
recall that $\pi$ is constant, according to~\eqref{Constants}, and recall that $z_2(k+1)-z_2(k)=T(\alpha(k)+\beta(k)u(k))$, we obtain the error dynamics:
\begin{eqnarray}
e_\pi(k+1)&=&G^a_T(e_\pi(k))=e_\pi(k)-T\gamma (k)\Phi(k) \Phi^T(k)e_\pi (k).\notag
\end{eqnarray}
This suggests using the Lyapunov function $E(e_\pi)=e_\pi^T e_\pi$ satisfying:
\begin{eqnarray}
E(G^a_T(e_\pi))-E(e_\pi)&=& e_\pi^T(I-T\gamma\,\Phi\Phi^T)(I-T\gamma\,\Phi\Phi^T)e_\pi-e_\pi^T e_\pi\notag\\
&=& -2T \gamma\, e_\pi^T\Phi\Phi^T e_\pi+T^2\gamma^2 e_\pi^T\Phi\Phi^T\Phi\Phi^Te_\pi\notag\\
&=& -2T\gamma\, e_\pi^T\Phi\Phi^T e_\pi+T^2\gamma^2 e_\pi^T\Phi(\Phi^T\Phi)\Phi^Te_\pi\notag\\
&=& -2T \gamma\, e_\pi^T\Phi\Phi^T e_\pi+T^2\gamma^2 \Phi^T\Phi e_\pi^T\Phi\Phi^T e_\pi\notag\\
&=& (-2T\gamma +T^2\gamma^2 \Phi^T\Phi)e_\pi^T\Phi\Phi^T e_\pi,\notag
\end{eqnarray}
where we used associativity of matrix multiplication and the fact that $\Phi^T\Phi$ is a scalar. We can now choose $\lambda_\pi\in \R^+$ and $T_0\in \R^+$ satisfying:
$$\lambda_\pi<2\underline{\gamma},\qquad T_0=\frac{-\lambda_\pi+2\underline{\gamma}}{\overline{\gamma}^2 N},$$
where $N$ is an upper bound for $\Phi^T\Phi$ whenever $u$ ranges in a compact set\footnote{The proof of the main result will show why we can assume the input to range on a compact set.}, to conclude that for all $T\in [0,T_0]$ we have:
$$E(G^a_T(e_\pi))-E(e_\pi)\le -\lambda_\pi T e_\pi^T\Phi\Phi^Te_\pi.$$
Only this inequality will play a role in the proof of the main results. Hence, any other observer leading to a similar inequality can be used.

Since the state $z$ is not available, we will use this observer with the estimate $\widehat{z}$ described in Section~\ref{Sec:StateEstimation}. Establishing that the observer still works correctly despite the use of the approximate model for the dynamics in its design and despite the use of $\widehat{z}$ will be done in the proof of the main result in Section~\ref{Sec:Main}. For the reader's convenience we provide here the observer equations to be used:
\begin{eqnarray}
\label{ObserverFinal1}
\widehat{\alpha}(k+1)&=&\widehat{\alpha}(k)-\gamma(k)T(\widehat{\alpha}(k)+u(k)\widehat{\beta}(k))+\gamma(k)(\widehat{z_2}(k+1)-\widehat{z_2}(k))\\
\label{ObserverFinal2}
\widehat{\beta}(k+1)&=&\widehat{\beta}(k)-\gamma(k)u(k)T(\widehat{\alpha}(k)+u(k)\widehat{\beta}(k))+\gamma(k)u(k)(\widehat{z_2}(k+1)-\widehat{z_2}(k)).
\end{eqnarray}

\section{Controller design}
\label{Sec:ControllerDesign}
If we assume the parameters $\alpha$ and $\beta$ to be known, we can design a family of controllers (parameterized by $T$) for the family of approximate models~\eqref{ModelOrder33a}-\eqref{ModelOrder33b} 
with the objective of  asymptotically stabilizing the origin in the following specific sense: there exists a symmetric and positive definite matrix $P$ and constants $\lambda_z,T_0\in \R^+$ so that $V(z)=z^TPz$ satisfies:
\begin{equation}
\label{DefineV}
V(z(k+1))-V(z(k))\le -\lambda_z T\Vert z(k)\Vert^2+O_{z^2}(T^2),
\end{equation}
for all $T$ in the interval $[0,T_0]$. The main result in this note will only require this inequality, hence, any family of controllers leading to~\eqref{DefineV} can be used. In this note we use the very simple family of controllers:
\begin{equation}
\label{Controller}
u=\beta^{-1}(-\alpha+Kz),
\end{equation}
where $K$ is a suitable matrix. Note that this family of controllers is, in fact, independent of $T$. To show that it suffices to consider this family, note that the approximate model~\eqref{ModelOrder33a}-\eqref{ModelOrder33b} can be written as:
$$z(k+1)=Az(k)+B\alpha(k)+B\beta(k)u(k),$$
where the matrices $A$ and $B$ are of the form:
$$A=I+A_1T,\qquad B=B_1T+B_2 T^2.$$
Since $(A_1,B_1)$ is a controllable pair, there exists a controller $u=Kz$ and a symmetric and positive definite matrix $P$ so that:
\begin{equation}
\label{DefineVAgain}
(A_1+B_1K)^TP+P(A_1+B_1K)= -Q,
\end{equation}
for some symmetric and positive definite matrix $Q$. Computing $V((A+BK)z)-V(z)$ provides:
\begin{eqnarray}
V((A+BK)z)-V(z) & = & z^T(I+(A_1+B_1K)T+B_2K T^2)^TP(I+(A_1+B_1K)T+B_2K T^2)z-z^TPz\notag\\
& = & z^T((A_1+B_1K)T)^TPz+z^TP((A_1+B_1K)T)z+O_{z^2}(T^2)+O_{z^2}(T^3)+O_{z^2}(T^4)\notag\\
&=& -Tz^TQz+O_{z^2}(T^2)\le -\lambda_{\min}(Q)T\Vert z\Vert^2+O_{z^2}(T^2),\notag
\end{eqnarray} 
which is the desired inequality~\eqref{DefineV}.

Since neither $\alpha$  and $\beta$ nor $z$ are exactly known, we use instead the control law:
\begin{equation}
\label{ControllerFinal}
u(k)=\widehat{\beta}^{-1}(k)\left(-\widehat{\alpha}(k)+K\widehat{z}(k)\right),
\end{equation}
obtained by replacing the unknown quantities by its estimates. This control law, in combination with the observer~\eqref{ObserverFinal1}-\eqref{ObserverFinal2}, defines a dynamic controller. The main result in the next section explains why such controller works despite being designed for an approximate model and assuming knowledge of the exact value of the parameters and state in its design.

\newpage
\section{Main result}
\label{Sec:Main}

\subsection{The noise-free scenario}

The first result shows that it is possible to use the controller with the parameter and state estimates so far described despite having been designed for an approximate model and assuming the exact knowledge of the parameters and state.

\begin{theorem}
\label{Theorem1}
Consider an unknown nonlinear system of the form~\eqref{UnknownSystem1a}-\eqref{UnknownSystem1b} where the output function $h$ has relative degree $2$ and $L_gL_fh$ is positive. In the absence of measurement noise, i.e., $d=0$, for any compact set $\mathcal{S}\subset \R^2$ of initial conditions containing the origin in its interior there exists a time $T^*\in \R^+$ and a constant $b\in \R^+$ (both depending on $\mathcal{S}$) so that for any sampling time $T\in [0,T^*]$ the controller~\eqref{ControllerFinal} using the parameter estimates provided by the observer~\eqref{ObserverFinal1}-\eqref{ObserverFinal2} and the state estimate~\eqref{StateEstimateFinal} renders the closed-loop trajectories bounded, i.e., $\Vert \widehat{\alpha}(k),\widehat{\beta}(k))\Vert\le b$ for all $k\in \N$, $\Vert x(t)\Vert\le b$ for all $t\in \R_0^+$ and $\lim_{t\to\infty} x(t)=0$.
\end{theorem}

The key technical ingredients of the proof are: 1) the Lyapunov analysis of observers designed for approximate models developed by Arcak and Nesic in~\cite{AN04}; 2) the Lyapunov analysis of controllers designed for approximate models developed by Nesic and Teel in~\cite{NT04}; 3) the Lyapunov analysis of Immersion and Invariance adaptive controllers presented by Astolfi and co-workers in~\cite{IIBook} even though we do not use Immersion and Invariance techniques in this note.

\begin{proof}
The proof will be based on the feedback linearized form~\eqref{UnknownSystem2a}-\eqref{UnknownSystem2c} of the dynamics rather than the original nonlinear form~\eqref{UnknownSystem1a}-\eqref{UnknownSystem1b}. This results in no loss of generality since both systems are related by the diffeomorphism $\Psi$. For simplicity, we will denote the set $\Psi(\mathcal{S})$ simply by $\mathcal{S}$. Since $\Psi$ is an homeomorphism, $\Psi(\mathcal{S})$ is still a compact set. In the same spirit, rather than working with the estimate $\widehat{\pi}$ we will work with $e_\pi$. We will see that boundedness of $e_\pi$ implies boundedness of $\widehat{\pi}$. 

\textbf{The initial transient:}
the state estimate $\widehat{z}$ requires $\rho$ samples to be collected while the input is being held constant. Let us denote by $u^*$ the fixed input that is used when the controller is used for the first two time steps, i.e., at times $k=0$ and $k=1$. This corresponds to an initial transient that must be analyzed separately. 

By applying Proposition~\ref{Prop:BoundTaylor} to the compact set $\mathcal{D}=\mathcal{S}\times\{u^*\}$ we conclude the existence of a time $2T_1$ so that trajectories are well defined for all $T\in [0,2T_1]$ and for all initial conditions in $\mathcal{S}$. We regard $2T_1$ as the time elapsed during the first two time steps under input $u^*$. The set of points reached under all these trajectories and for all $T\in [0,2T_1]$ is denoted by $Z$. At time step $k=2$ both state and parameter estimates are already available (recall that observer~\eqref{ObserverFinal1}-\eqref{ObserverFinal2} requires $\widehat{z}_2(2)-\widehat{z}_2(1)$ to produce the estimate $\widehat{\pi}(2)$). Let us denote by $E$ the set of possible parameter estimation errors $e_\pi=\pi-\widehat{\pi}$ at time $k=2$ depending on the different initial conditions. Since the estimates $\widehat{\pi}$ are continuous functions of the initial condition $z(0)$ that belongs to the compact set $\mathcal{S}$, $E$ is a bounded set. We now consider the set $R$ defined as the smallest sublevel set of $V+W$ that contains $Z\times E$ where $W$ is defined by $W(e_\pi)=e_\pi^T e_\pi$ and $V$ is the Lyapunov satisfying~\eqref{DefineV}. Our objective is to show that $R$ is an invariant set. 
 
 \textbf{Existence of solutions one step beyond the transient:} we first show that it is possible to continue the solutions from $R$ by employing again Proposition~\ref{Prop:BoundTaylor}. For future use, we define the projections $\pi_1:\R^2\to\R$, $\pi_Z:\R^2\times\R^2\to \R^2$, and $\pi_E:\R^2\times\R^2\to \R^2$  defined by $\pi_1(z_1,z_2)=z_1$, $\pi_Z(z,e_\pi)=z$, and $\pi_E(z,e_\pi)=e_\pi$. The controller~\eqref{ControllerFinal} is a function of $\widehat{z}$ and $\widehat{\pi}$. However, since $\widehat{\pi}=\pi-e_\pi$ and both $\pi$ and $\widehat{z}$ are functions of $z$, we can regard the controller as a smooth function of $z$ and $e_\pi$. We can thus consider the set of inputs $U\subset \R$ defined by all the inputs obtained via~\eqref{ControllerFinal} when $z$ ranges in $\pi_Z(R)$, $e_\pi$ ranges in $\pi_E(R)$ and $\widehat{z}$ is given by~\eqref{StateEstimateFinal} with $Y(\rho)$ ranging in $\left(\pi_1\circ \pi_Z(R)\right)^{2\rho}$, the $2\rho$-fold Cartesian product of $\pi_1\circ\pi_Z(R)$. By taking its closure, if needed, we can assume the set $\pi_Z(R)\times U$ to be compact  and apply Proposition~\ref{Prop:BoundTaylor} to obtain a time $T_2$ ensuring that solutions starting at $\pi_Z(R)$ exist for all $T\in [0,T_2]$. Moreover, Proposition~\ref{Prop:BoundTaylor} ensures the existence of a constant $M$ for which the bound~\eqref{BoundInProp} holds and, as a consequence, the approximate model~\eqref{ModelOrder33a}-\eqref{ModelOrder33b} is valid for any solution with initial condition in $R$ and any input in $U$. If $T_1<T_2$ we proceed by only considering sampling times in $[0,T_1]$ and note that none of the conclusions reached so far changes. If $T_1>T_2$, we can use sampling times in $[0,T_2]$ while noting that all the reached conclusions remain valid by redefining $Z$ to be the set of points reached for any time in $[0,T_2]$ (if the conclusions holds for the (non-strictly) larger $Z$ set they also hold for the (non-strictly) smaller $Z$ obtained by reducing $T_1$ to $T_2$).

\textbf{Invariance of the set $R$:} we can now establish invariance of $R$ by computing $W(F_T^e(z,u),G_T^e(e_\pi))-W(z,e_\pi)$ for all $(z,e_\pi)\in R$ in several steps.

In the first step we establish that the evolution of $V$, under $F_T^e$ and the controller~\eqref{ControllerFinal}, equals the evolution of $V$ under $F_T^a$ and the controller~\eqref{ControllerFinal} up to $O(T^2)$ terms:
\begin{eqnarray}
V(F_T^e(z,u))-V(z)&=& V(F_T^a(z,u))-V(z)+V(F_T^e(z,u))-V(F_T^a(z,u))\notag\\
&\le& V(F_T^a(z,u))-V(z)+\Vert V(F_T^e(z,u))-V(F_T^a(z,u))\Vert\notag\\
\label{NormToSquaredNorm}
&\le& V(F_T^a(z,u))-V(z)+c\Vert F_T^e(z,u)-F_T^a(z,u)\Vert^2\notag\\
\label{LyapunovIneqV}
&\le& V(F_T^a(z,u))-V(z)+cM^2{T}^6\Vert (z,u-u_0)\Vert^2,
\end{eqnarray}
for some constant $c\in \R^+$ and where the second inequality is proved in the Appendix and the third follows from Proposition~\ref{Prop:BoundTaylor}. We now consider the term $\Vert (z,u-u_0)\Vert$ in more detail:
\begin{eqnarray}
\Vert (z,u-u_0)\Vert &\leq & \Vert z \Vert + \left\Vert \widehat{\beta}^{-1}(-\widehat{\alpha}+K\widehat{z})-u_0 \right\Vert\notag\\
&= &\Vert z \Vert + \left\Vert \frac{1}{\beta-e_\beta}(-\widehat{\alpha}+K\widehat{z})-u_0 \right\Vert  \notag\\
&= &\Vert z \Vert + \left\Vert \left(\frac{1}{\beta}+\frac{e_\beta}{\beta(\beta-e_\beta)}\right)(-\widehat{\alpha}+K\widehat{z})-u_0 \right\Vert \notag \\
&= &\Vert z \Vert + \left\Vert \left(\frac{1}{\beta}+\frac{e_\beta}{\beta\widehat{\beta}}\right)(-\widehat{\alpha}+K\widehat{z})-u_0\right \Vert\notag \\
&\leq &\Vert z \Vert + \left\Vert \beta^{-1}(-\alpha+Kz)-u_0\right\Vert+\left\Vert \beta^{-1}(e_\alpha-Ke_z)+\beta^{-1}e_\beta\widehat{\beta}^{-1}(-\widehat{\alpha}+K\widehat{z})\right \Vert \notag \\
&\leq &\Vert z \Vert + \left\Vert \beta^{-1}(-\alpha+Kz)-u_0\right\Vert+\left\Vert -\beta^{-1}Ke_z+\beta^{-1}e_\alpha+\beta^{-1}e_\beta u\right \Vert \notag \\
\label{HandlingInput1}
&\leq &\Vert z \Vert + \left\Vert \beta^{-1}(-\alpha+Kz)-u_0\right\Vert+\left\Vert \beta^{-1} Ke_z \right\Vert + \left\Vert \beta^{-1}\Phi^T e_\pi \right\Vert .
\end{eqnarray}
Noting the function $\beta^{-1}(z)(-\alpha(z)+Kz)$ is Lipschitz continuous (with constant $L$) on $\pi_Z(R)$, and that it produces the value $u_0$ at $z=0$, we conclude that: 
\begin{equation}
\label{HandlingInput2}
\left\Vert\beta^{-1}\left(-\alpha+Kz\right)-u_0\right\Vert\le L\Vert z\Vert.
\end{equation}
By combining the previous bounds~\eqref{HandlingInput1} and~\eqref{HandlingInput2} with~\eqref{LyapunovIneqV} we obtain:
\begin{eqnarray}
V(F_T^e(z,u))-V(z)&\le& V(F_T^a(z,u))-V(z)+cdM^2{T}^6\left(\Vert z\Vert^2+\Vert e_z\Vert^2+\Vert\Phi^Te_{\pi}\Vert^2\right),\\
\label{FinalIneq1}
&=& V(F_T^a(z,u))-V(z)+O_{z^2}(T^6)+O_{e_z^2}(T^6)+O_{(\Phi^T e_\pi)^2}(T^6)\\
\label{Step1}
&=& V(F_T^a(z,u))-V(z)+O_{z^2}(T^2)+O_{(\Phi^T e_\pi)^2}(T^2),
\end{eqnarray}
where $d\in \R^+$ is a suitable constant and we used the fact that $\beta^{-1}$ is bounded on the compact $\pi_Z(R)$ and equality~\eqref{StateEstimationError} combined with the previous argument to replace $O_{e_z^2}(T^6)$ with $O_{z^2}(T^6)$. Note that, in particular, we established that $O_{(z,u-u_0)}(T)$ can be replaced with $O_z(T)+O_{\Phi^T e_\pi}(T)$. We will be using this fact in the remainder of the proof without explicitly acknowledging it.

In the second step we show that $V(F_T^a(z,u))-V(z)$, when using the controller~\eqref{ControllerFinal}, is negative definite up to $O(T^2)$ terms. The dynamics is given by:
\begin{eqnarray}
F^a_T(z,u)&=& Az +B\alpha +B\beta \widehat{\beta}^{-1} \left(-\widehat{\alpha} +K\widehat{z} \right) \notag \\
&=&Az +B\alpha +B\left(\widehat{\beta} +e_\beta \right)\widehat{\beta}^{-1} \left(-\widehat{\alpha} +K\widehat{z} \right) \notag \\
&=&Az +B\alpha +B\left(-\widehat{\alpha} +K\widehat{z} \right)+B e_\beta \widehat{\beta}^{-1} \left(-\widehat{\alpha} +K\widehat{z} \right) \notag \\
&=&Az +B\alpha +B\left(-\alpha +e_\alpha +Kz -Ke_z \right)+B e_\beta \widehat{\beta}^{-1} \left(-\widehat{\alpha} +K\widehat{z} \right) \notag \\
&=&Az +BKz -BKe_z +Be_\alpha +B e_\beta \widehat{\beta}^{-1} \left(-\widehat{\alpha} +K\widehat{z} \right) \notag \\
&=&Az +BKz -BKe_z +Be_\alpha +B e_\beta u  \notag \\
&=&(A+BK)z -BKe_z +B\Phi^Te_\pi . \notag
\end{eqnarray}
By interpreting the resulting system as a linear asymptotically stable system perturbed by $-BKe_z+B\Phi^Te_\pi$, and by invoking ISS, we conclude the existence of constants $\lambda_z, \omega\in \R^+$ so that the following inequality holds:
$$V(F_T^a(z,u))-V(z)\le -\lambda_z T\Vert z\Vert^2+\omega e_z^TK^T B^TBKe_z+\omega e_\pi^T\Phi B^T B\Phi^Te_\pi.$$
Recalling that~\eqref{StateEstimationError} shows that we can replace $O_{e_z}$ with $O_{(z,u-u_0)}$, and since $B^TB=T^2+T^4/4$, we have:
\begin{eqnarray}
\label{Step2}
V(F_T^a(z,u))-V(z)&\le& -\lambda_z T\Vert z\Vert^2+O_{z^2}(T^2)+O_{(\Phi^Te_\pi)^2}(T^2).
\end{eqnarray}

In the third step we analyze the effect of using the estimate $\widehat{z}$ when implementing the observer~\eqref{ObserverFinal1}-\eqref{ObserverFinal2} and also the effect of using the approximate model $G^a_T$, instead of the exact model $G^e_T$, when designing the observer. By redoing the analysis in Section~\ref{SecParameter} with the parameter estimation error defined as in~\eqref{ParameterError}, using the observer~\eqref{ObserverFinal1}-\eqref{ObserverFinal2}, and noting that $\widehat{z_2}$ satisfies~\eqref{ModelOrder33b}, we obtain:
$$E(G_T^e(e_\pi))-E(e_\pi)\le -\lambda_\pi T e_\pi^T\Phi\Phi^Te_\pi+\mathcal{E}^T\mathcal{E},$$
where $\mathcal{E}$ is given by:
$$\mathcal{E}(k)=\gamma(k)\Phi(k)O_{(z,u-u_0)}(T).$$ 
In view of $\gamma\Phi$ being bounded on the compact $R$, we can replace $\gamma(k)\Phi(k)O_{(z,u-u_0)}(T)$ with $O_{z}(T)+O_{\Phi^T e_\pi}(T)$ to obtain:
\begin{equation}
\label{Step3}
E(G_T^e(e_\pi))-E(e_\pi)\le -\lambda_\pi T e_\pi^T\Phi\Phi^Te_\pi+O_{z^2}(T^2)+O_{(\Phi^T e_\pi)^2}(T^2).
\end{equation}

We now put the three intermediate steps,~\eqref{Step1},~\eqref{Step2}, and~\eqref{Step3}, together:
\begin{eqnarray}
V(F_T^e(z,u))+W(G_T^e(e_\pi))-V(z)-W(e_\pi)&\le& -\lambda_z T\Vert z\Vert^2-\lambda_\pi T\Vert \Phi^Te_\pi\Vert^2+O_{z^2}(T^2)+O_{(\Phi^T e_\pi)^2}(T^2)\notag\\
&\le & -\lambda_z T\Vert z\Vert^2-\lambda_\pi T\Vert \Phi^Te_\pi\Vert^2+M_1T^2\Vert z\Vert^2+M_2T^2\Vert \Phi^T e_\pi\Vert^2,\notag
\end{eqnarray}
where $M_1,M_2\in\R^+$ are constants following from the definition of the $O(T^2)$ terms. If we choose $\lambda\in \R^+$ and $T_3\in \R^+$ satisfying:
$$\lambda<\min\{\lambda_z,\lambda_{\pi}\},\qquad T_3=\frac{-\lambda+\min\{\lambda_z,\lambda_{\pi}\}}{\max\{M_1,M_2\}},$$
it follows that for all $T\in [0,T_3]$ we have:
\begin{equation}
\label{LastInequality}
V(F_T^e(z,u))+W(G_T^e(e_\pi,t))-V(z)-W(e_\pi)\le -\lambda T\Vert z\Vert^2-\lambda T\Vert \Phi^Te_\pi\Vert^2.
\end{equation}
Therefore, for any $T\in [0,T_4]$, $T_4=\min\{T_1,T_2,T_3\}$, we have that $R$ remains invariant. By noting that trajectories remain in $R$ for any time in $[0,T_4]$ we conclude that we can apply the same argument to establish that trajectories remain in $R$ for any consecutive time step since we only assumed that inputs were generated based on samples of $z_1$ that remained in $\pi_Z\circ \pi_1(R)$. Compactness of $R$ establishes that trajectories are bounded and thus there exists a constant $b_1\in \R^+$ so that $\Vert e_\pi(k)\Vert\le b_1$ and $\Vert z(k)\Vert \le b_1$ for all $k\in \N$. Since $\pi$ is a smooth function of $z$ and $z$ is bounded, boundedness of $e_\pi=\pi-\widehat{\pi}$ implies boundedness of the estimate $\widehat{\pi}$, i.e., there exists a constant $b_2\in \R^+$ so that $\Vert \widehat{\pi}(k)\Vert\le b_2$ for all $k\in \N$. Moreover,~\eqref{LastInequality} enables the use of LaSalle's invariance principle to conclude that trajectories will converge to the largest invariant set contained in the set defined by the equality:
$$\Vert z\Vert^2+ \Vert \Phi^Te_\pi\Vert^2=0.$$
In particular, $z$ will converge to the origin. Invoking Theorem~1 in~\cite{NTS99}, combined with invariance of $R$ and smoothness of the dynamics, we conclude that the solutions of~\eqref{UnknownSystem1a} when using the controller~\eqref{ControllerFinal} with the parameter estimates provided by the observer~\eqref{ObserverFinal1}-\eqref{ObserverFinal2} and the state estimate~\eqref{StateEstimateFinal}, are bounded, i.e., there exists a constant $b_3\in \R^+$ so that $\Vert x(t)\Vert\le b_3$ and, moreover,  $\lim_{t\to \infty}x(t)=0$. Hence, by taking $b=\max\{b_1,b_2,b_3\}$ we conclude the proof.
\end{proof}

As is typical in adaptive control there is no guarantee the parameter estimates converge to the true values although $\Phi^T e_\pi$ converges to zero. A suitable persistency of excitation assumption on $\Phi$ can be derived so that convergence of $\Phi^T e_\pi$ to zero implies convergence of $e_\pi$ to zero. 

\subsection{The noisy scenario}

As is known from adaptive control, e.g.,~\cite{RobustAdaptiveControlBook,StableAdaptiveSystemsBook}, this type of controllers is not robust to measurement errors in the sense that it is not possible to guarantee boundedness of the parameter estimates. Hence, we now use the stronger assumption that bounds for $\beta$ are known and modify the parameter estimator by projecting the estimate for $\beta$ on the set $\left[\underline{\beta},\overline{\beta}\right]$. This is achieved by replacing~\eqref{ObserverFinal2} with:
\begin{equation}
\label{ProjectedObserver}
\widehat{\beta}(k+1)=P\left(\widehat{\beta}(k)-\gamma(k)u(k)T(\widehat{\alpha}(k)+u(k)\widehat{\beta}(k))+\gamma(k)u(k)(\widehat{z_2}(k+1)-\widehat{z_2}(k))\right).
\end{equation}
where $P$ is the projection defined by:
$$P(r)=\left\{\begin{array}{ccc}
\underline{\beta} & \text{if}&r<\underline{\beta}\\
r & \text{if}& \underline{\beta}\le r\le\overline{\beta} \\
\overline{\beta} & \text{if}&r>\overline{\beta}.
\end{array}\right.$$
The following sequence of inequalities:
$$
\vert e_\beta\vert  =  \left\vert \beta-P(\widehat{\beta})\right\vert  =  \left\vert P(\beta)-P(\widehat{\beta})\right\vert\notag \le \left\vert \beta-\widehat{\beta}\right\vert,\notag
$$
which is a consequence of  convexity of $\left[\underline{\beta},\overline{\beta}\right]$ implying $\vert P(r)-P(s)\vert\le \vert r-s\vert$, shows that by replacing~\eqref{ObserverFinal2} with~\eqref{ProjectedObserver} we do not alter the conclusions of Theorem~\ref{Theorem1} in the absence of measurement noise. In the presence of such noise, the projection $P$ enforces boundedness of the estimate of $\beta$. Moreover, the state estimate $\widehat{z}$, given by~\eqref{StateEstimateFinal}, is also guaranteed to be bounded for bounded measurement noise in virtue of being given by a linear map of the observations. However, revising the argument in the proof of Theorem~\ref{Theorem1} to account for a bounded error in $\widehat{z}$ with bound $r$, ie.., $\Vert e_z\Vert\le r$,  results in~\eqref{LastInequality} becoming:
\begin{equation}
\label{LastIneqError}
V(F_T^e(z,u))+W(G_T^e(e_\pi,t))-V(z)-W(e_\pi)\le -\lambda T\Vert z\Vert^2-\lambda T\Vert \Phi^Te_\pi\Vert^2+c r^2
\end{equation}
for a suitable constant $c\in \R^+$. The right-hand side of~\eqref{LastIneqError} is zero when:
\begin{equation}
\label{Zero}
\Vert z\Vert^2+\Vert\Phi^T e_\pi\Vert^2=\frac{c}{\lambda T}r^2.
\end{equation}
Unfortunately, since $\Phi\Phi^T$ is not full rank, the previous equality does not define a compact set for $(z,e_\pi)$. However, since $\Phi\Phi^T$ has rank $1$ and $e_\beta$ is bounded, the set defined by:
$$\left\{\begin{array}{l}\Vert z\Vert^2+\Vert\Phi^T e_\pi\Vert^2=\frac{c}{\lambda T}r^2\\
\vert e_\beta\vert\le 2 \left(\overline{\beta}-\underline{\beta}\right)\end{array}\right. ,$$
is indeed bounded. This can be seen by noticing that equality~\eqref{Zero} implies $\Vert z\Vert^2\le \frac{c}{\lambda T}r^2$ showing that $z$ is bounded. Moreover, equality~\eqref{Zero} also implies $e_\pi^T\Phi\Phi^T e_\pi=\Vert\Phi^T e_\pi\Vert^2\le\frac{c}{\lambda T}r^2=c_1$. Since the inequality $\vert e_\beta\vert\le 2 \left(\overline{\beta}-\underline{\beta}\right)$ can be cast as:
$$e_\pi^T Qe_\pi=e_\beta^2\le 4 \left(\overline{\beta}-\underline{\beta}\right)^2=c_2,\qquad Q=\begin{bmatrix}0&0\\0&1\end{bmatrix},$$
we conclude:
$$e_\pi^T(\Phi\Phi^T+Q)e_\pi\le c_1+c_2.$$
It can now easily be checked that $\Phi\Phi^T+Q$ is positive definite thereby showing that $e_\pi$ is bounded. We summarize this discussion in the following result.

\begin{theorem}
\label{Theorem2}
Consider an unknown nonlinear system of the form~\eqref{UnknownSystem1a}-\eqref{UnknownSystem1b} where the output function $h$ has relative degree $2$ and  assume the existence of two constants $\underline{\beta},\overline{\beta}\in \R^+$ satisfying $\underline{\beta}\le L_gL_fh(x)\le \overline{\beta}$ for all $x\in \R^2$.  For any compact set $\mathcal{S}\subset \R^2$ of initial conditions containing the origin in its interior there exists a time $T^*\in \R^+$ and a constant $b\in \R^+$ (both depending on $\mathcal{S}$) so that for any sampling time $T\in [0,T^*]$ the controller~\eqref{ControllerFinal} using the parameter estimates provided by the observer~\eqref{ObserverFinal1}-\eqref{ProjectedObserver} and the state estimate~\eqref{StateEstimateFinal} renders the closed-loop trajectories bounded, i.e., $\Vert \widehat{\alpha}(k),\widehat{\beta}(k))\Vert\le b$ for all $k\in \N$, $\Vert x(t)\Vert\le b$ for all $t\in \R_0^+$ and $x(t)$ converges to a ball around the origin of size proportional to the magnitude of the measurement noise. In particular, in the absence of measurement noise, i.e., $d=0$, we have $\lim_{t\to\infty} x(t)=0$.
\end{theorem}

\section*{Appendix}
\label{Sec:Appendix}

The inequality $\vert V(a)-V(b)\vert\le c\Vert a-b\Vert^2$ is equivalent to (we assume $V(a)>V(b)$ without loss of generality):
$$\sup_{a,b}\phi(a,b)=\sup_{a,b}\frac{V(a)-V(b)}{(a-b)^T(a-b)}\le c.$$
We fist note that $\phi(a,b)$ is homogeneous, i.e., $\phi(\delta a,\delta b)=\phi(a,b)$ for any $\delta\in \R$. Hence, it suffices to compute the supremum for vectors $(a,b)$ satisfying $(a-b)^T(a-b)=1$, i.e., it suffices to compute $\sup_{a,b} V(a)-V(b)$. Since $V(a)-V(b)$ is a continuous function and $a,b$ range in a compact set, a finite upper bound exists.

\bibliographystyle{alpha}
\bibliography{arXivFinal}
\end{document}